\documentclass[11pt]{article} %for pdfLaTeX
\usepackage{cite}
\usepackage{slashed}
\usepackage{latexsym}
\usepackage{amssymb}
\usepackage{amsmath}
\usepackage{hyperref}
\hypersetup{colorlinks=true, allcolors=blue}
\usepackage{graphicx}% Include figure files

\usepackage{slashed}
\usepackage{mathbbol}

\usepackage[title]{appendix}

\usepackage{color}
\begin{document}
\pagestyle{empty}

\begin{flushright}
KEK--TH--2462
\end{flushright}

\vspace{3cm}

\begin{center}

{\bf\LARGE  
Precision $\mu^+\mu^+$ and $\mu^+e^-$ elastic scatterings } \\

\vspace*{1.5cm}
{\large 
Yu Hamada$^{1}$, Ryuichiro Kitano$^{1,2}$, Ryutaro Matsudo$^{1,3}$,
Hiromasa Takaura$^{1}$
} \\
\vspace*{0.5cm}

{\it 
$^1$KEK Theory Center, Tsukuba 305-0801,
Japan\\
$^2$Graduate University for Advanced Studies (Sokendai), Tsukuba
305-0801, Japan\\
$^3$Department of Physics, Taiwan University, Taipei 10617, Taiwan
}

\end{center}

\vspace*{1.0cm}

\begin{abstract}
{\normalsize
Expected precisions of measurements of the elastic scattering cross
sections are estimated for $\mu^+ \mu^+$ and $\mu^+ e^-$
colliders, which are recently proposed as future realistic possibilities ($\mu$TRISTAN).
Comparing with contributions from possible new physics
represented by higher dimensional operators, we find that the
measurements at a TeV energy $\mu^+\mu^+$ collider can probe the scale
of new physics up to $O(100)$~TeV. A $\mu^+ e^-$ collider for the
Higgs boson factory can also improve the electroweak precision test. 
}
\end{abstract} 

%%%%%%%%%%%%%%%%%%%%%%%%%%%%%%%%%%%%%%%%%%%%%%%%%%%%%%%%%%%%%%%%%%%%%%%%%%%%
\newpage
\baselineskip=18pt
\setcounter{page}{2}
\pagestyle{plain}

\setcounter{footnote}{0}

\tableofcontents
\noindent\hrulefill

\section{Introduction}

The technology of particle accelerations has made impressive progress
since Van de Graaff and Cockcroft-Walton in the early 20th century.
The energy frontier has now reached
the center of mass energy of $O(10)$~TeV for hadron collisions,
but still even higher energies are demanded
from particle physics.

One of the important tasks in the future accelerator experiments will
be to gather information of new physics that enables us to build a
guidance for the big picture.
In particular, at lepton colliders, well-defined initial states provide
clean environment for the precision tests of scattering processes.
Indeed, $e^+ e^-$ colliders such as the LEP experiments at CERN have
given important inputs towards the complete picture of the Standard
Model~\cite{ALEPH:2005ab,ALEPH:2010aa,ALEPH:2013dgf}.
In addition, the precision tests of the scattering processes have been
giving severe constraints on the new physics contributions~\cite{Derman:1979zc,Czarnecki:1995fw,Ramsey-Musolf:1999qyv,Czarnecki:2000ic,SLACE158:2005uay,Kumar:2013yoa,MOLLER:2014iki}.

After the establishment of the Standard Model, at least of its
particle content and its gauge group, one can now parametrize the
effects of physics beyond the Standard Model as coefficients of higher
dimensional operators in the Lagrangian of the Standard Model effective field theory (SMEFT)~\cite{Buchmuller:1985jz,Grzadkowski:2010es,Brivio:2017vri,Ellis:2020unq}. Although there are
possibilities of having light new particles, new physics models
motivated by the UV origin of the Higgs boson can be generically fall
into this type where new particles and/or new interactions appear at
TeV or multi-TeV scale.
Once a new lepton collider is to be built, the simplest and also one of
the most important questions is how well such coefficients can be
measured.
At lepton colliders, there are enormous amount of elastic scattering
events, which typically have a sharp peak in the forward region. The new
physics effects interfere with the Standard-Model amplitude and modify
the angular distribution mostly in the central region. Such anomalous
distributions can be detected at a high accuracy and thus provides
good probe of microscopic physics.
Indeed, prospects at the ILC experiments have been studied, and significant improvement over the current limits has been reported~\cite{Behnke:2013xla, Ellis:2015sca}.

Recently, a new collider experiment, $\mu$TRISTAN~\cite{Hamada:2022mua}, using the
ultra-cold muon~\cite{Kondo:2018rzx}, has been proposed. The technology of $\mu^+$ cooling by the laser ionization of the muonium (the $\mu^+ e^-$ bound
state) has been developed for the muon $g-2$/EDM experiment at J-PARC~\cite{Abe:2019thb}.
This technology enables us to consider realistic $\mu^+ e^-$ and $\mu^+ \mu^+$ colliders. 
As a possible design for a 3~km storage ring, it has been
considered to accelerate $\mu^+$ and $e^-$ beams up to 1~TeV and
30~GeV, respectively, with which the $\mu^+ e^-$ and $\mu^+ \mu^+$
colliders with the center-of-mass energy to be 346~GeV and 2~TeV can be
realized.
The luminosities are estimated to be at the level of
$10^{33-34}$~cm$^{-2}$~s$^{-1}$, with which the $\mu^+ e^-$ collider can
be a good Higgs boson factory whereas direct new particle searches are
possible at the high energy $\mu^+ \mu^+$ collider.

In this paper, we examine how well $\mu$TRISTAN can measure the
elastic $\mu^+e^-$ and $\mu^+ \mu^+$ scattering processes.
Since the contributions from, for example, four-fermion operators are
larger for higher energies, the $\mu^+ \mu^+$ collider at 2~TeV should
give the most stringent constraints on the coefficients. We indeed
find that the collider can probe the energy scale of $O(100)$~TeV.
We include all the SMEFT dimension-six operators that contribute to the
elastic scattering. We find that the $\mu^+ e^-$ collider at 346~GeV can
improve the constraints on the electroweak precision observables such
as $S$ and $T$ parameters~\cite{Peskin:1990zt,Altarelli:1990zd}.

This paper is organized as follows. 
In Sec.~\ref{sec:2}, we give a review on
SMEFT and
specify the operators that we consider.
We give the expected
constraints on SMEFT operators 
at a $\mu^+ \mu^+$ collider
(Sec.~\ref{sec:3}) and at an $e^- \mu^+$
collider (Sec.~\ref{sec:4}).
We compare our result with the current
constraints in Sec.~\ref{sec:5}.
Sec.~\ref{sec:6} is devoted to the
summary.
In App.~\ref{app:A}, we present
kinematic formulas relevant to an $e^- \mu^+$
collider, whose beam energies are
asymmetric.

%...
\section{SMEFT Lagrangian}
\label{sec:2}
In this section, we present
necessary formulas 
to calculate scatterings 
$e^- \mu^+ \to e^- \mu^+$
and
$\mu^+ \mu^+ \to \mu^+ \mu^+$ 
within SMEFT. 
This includes a review on the relations between $(\alpha, M_Z, G_F)$
and SMEFT Lagrangian parameters.
We consider up to dimension-six operators 
adopting the basis of ref.~\cite{Alonso:2013hga}. 
All calculations in this paper are performed at the tree level.

\subsection{Gauge fields}
Let us first consider Lagrangian for the gauge fields. The following
four higher dimensional operators
\begin{align}
&Q_{HW}=H^{\dagger} H W_{\mu \nu}^I W^{I \mu \nu} \\    
&Q_{HB}=H^{\dagger} H B_{\mu \nu} B^{\mu \nu}     \\
&Q_{HWB}=H^{\dagger} \tau^I H W^I_{\mu \nu} B^{\mu \nu} \\
&Q_{HD}=(H^{\dagger} D_{\mu} H)^* (H^{\dagger} D_{\mu} H)
\end{align}
are involved. The Lagrangian is given by
\begin{align}
\mathcal{L}_{\rm gauge}
&=-\frac{1}{4} B^{\mu \nu} B_{\mu \nu}-\frac{1}{4} W_{\mu \nu}^I W^{I \mu \nu}+D_{\mu} H^{\dagger} D^{\mu} H  \nonumber \\  
&\qquad{}+C_{HW} H^{\dagger} H W_{\mu \nu}^I W^{I \mu \nu}+C_{HB} H^{\dagger} H B_{\mu \nu} B^{\mu \nu}
+C_{HWB} H^{\dagger} \tau^I H W^I_{\mu \nu} B^{\mu \nu}\notag\\
&\qquad+C_{HD} (H^{\dagger} D_{\mu} H)^* (H^{\dagger} D_{\mu} H) \nonumber \\ 
&=
-\frac{1}{4} (1-2 v^2 C_{HB})B^{\mu \nu} B_{\mu \nu}
-\frac{1}{4} (1-2 v^2 C_{HW}) W_{\mu \nu}^3 W^{3 \mu \nu}
-\frac{v^2}{2} C_{HWB} W_{\mu \nu}^3 B^{\mu \nu} \nonumber \\ 
&\qquad{}+(W^3_{\mu}, B_{\mu}) M
\begin{pmatrix}
W^3_{\mu} \\
B_{\mu}
\end{pmatrix}
-\frac{1}{2}  (1-2 v^2 C_{HW}) W_{\mu \nu}^+ W^{\mu \nu -}
+\frac{g^2 v^2}{4} W_{\mu}^+ W^{\mu-}
+\cdots .
\end{align}
Here we define the matrix $M$ by
\begin{align}
M
&=\left(1+\frac{v^2}{2} C_{HD} \right)
\begin{pmatrix}
\frac{g^2 v^2}{8} & -\frac{g {g'} v^2}{8} \\
 -\frac{g {g'} v^2}{8} &  \frac{{g'}^2 v^2}{8}
\end{pmatrix} ,
\end{align}
and define $W^+=\frac{1}{\sqrt{2}}(W^1-i W^2)$ and
$W^-=\frac{1}{\sqrt{2}}(W^1+i W^2)$.
In the final equality, we show explicitly only the part where 
the Higgs VEV is substituted and quadratic part in the gauge fields.
We use  $D_{\mu} H=(\partial_{\mu}+i \frac{g}{2} \tau^I W^I_{\mu}+i \frac{{g'}}{2} B_{\mu}) H$
and 
$\langle H \rangle=\frac{1}{\sqrt{2}} \begin{pmatrix} 0 \\ v \end{pmatrix}$.

Let us consider $A_{\mu}$ and $Z_{\mu}$ part, i.e., $B_{\mu}$ and $W_{\mu}^3$ part. We can diagonalize the matrix $M$ by
\begin{align}
\begin{pmatrix}
W^3_{\mu} \\
B_{\mu}
\end{pmatrix}
=
\begin{pmatrix}
\cos{\theta_W} & \sin{\theta_W} \\
-\sin{\theta_W} & \cos{\theta_W} 
\end{pmatrix}
\begin{pmatrix}
Z_{\mu} \\
A_{\mu}
\end{pmatrix} ,
\end{align}
where $\cos{\theta_W}=g/\sqrt{g^2+{g'}^2}$ and $\sin{\theta_W}={g'}/\sqrt{g^2+{g'}^2}$. Note that $\theta_W$ differs from
the SM value of the Weinberg angle. This is because $g$ and ${g'}$ are {\it{different from their SM values}}, which are determined by assuming the SM.
The mass term turns into
\begin{equation}
(W^3_{\mu}, B_{\mu}) M
\begin{pmatrix}
W^3_{\mu} \\
B_{\mu}
\end{pmatrix}
=\frac{1}{2} (1+\frac{v^2}{2} C_{HD}) \frac{1}{4} (g^2+{g'}^2) v^2 Z_{\mu} Z^{\mu} .
\end{equation}
Then the kinetic terms are written in the form of
\begin{equation}
\mathcal{L}
=-\frac{1}{4} (1+\epsilon_1) F_{\mu \nu} F^{\mu \nu}-\frac{1}{4} (1+\epsilon_2) Z_{\mu \nu} Z^{\mu \nu}
+\epsilon_3 F_{\mu \nu} Z^{\mu \nu} .
\end{equation}
Here the $\epsilon_i$ parameters are given by
\begin{equation}
\epsilon_1=
- 2 v^2 C_{HB} \cos^2{\theta_W}-2 v^2 C_{HW} \sin^2{\theta_W} 
+2 v^2 C_{HWB} \cos{\theta_W} \sin{\theta_W} ,
\end{equation}
\begin{equation}
\epsilon_2
=-2 v^2 C_{HB} \sin^2{\theta_W} -2 v^2 C_{HW} \cos^2{\theta_W}
-2 v^2 C_{HWB} \cos{\theta_W} \sin{\theta_W},
\end{equation}
\begin{equation}
\epsilon_3
=-v^2 C_{HB} \cos{\theta_W} \sin{\theta_W} +v^2 C_{HW} \cos{\theta_W} \sin{\theta_W} 
-\frac{v^2}{2} C_{HWB} (\cos^2{\theta_W}-\sin^2{\theta_W}) .
\end{equation}

After the following (non-orthogonal) transformation
\begin{equation}
\begin{cases}
A_{\mu} \to A_{\mu}+ 2 \epsilon_3 Z_{\mu} \\
Z_{\mu} \to Z_{\mu}
\end{cases}  ,
\end{equation}
the kinetic term becomes
\begin{equation}
\mathcal{L}
=-\frac{1}{4} (1+\epsilon_1) F_{\mu \nu} F^{\mu \nu}
-\frac{1}{4} (1+\epsilon_2) Z_{\mu \nu} Z^{\mu \nu} ,
\end{equation}
where the kinetic mixing is eliminated.
Note that the mass term is invariant under the above transformation
because we shift only the zero-mass field.
To make the kinetic terms canonical, we need further rescaling by $\sqrt{1+\epsilon_1}$ and $\sqrt{1+\epsilon_2}$ for $A_\mu$ and $Z_\mu$, respectively.
As a whole, we should use the following correspondence:
\begin{equation}
\begin{pmatrix}
W^3_{\mu} \\
B_{\mu}
\end{pmatrix}
=
\begin{pmatrix}
\cos{\theta_W} & \sin{\theta_W} \\
-\sin{\theta_W} & \cos{\theta_W} 
\end{pmatrix}
\begin{pmatrix}
\frac{1}{\sqrt{1+\epsilon_2}} Z_{\mu} \\
\frac{1}{\sqrt{1+\epsilon_1}} A_{\mu}+2 \epsilon_3 Z_{\mu}
\end{pmatrix} .
\end{equation}
Under this understanding the quadratic term of gauge fields becomes
\begin{equation}
\mathcal{L}
=-\frac{1}{4} F_{\mu \nu} F^{\mu \nu}
-\frac{1}{4}  Z_{\mu \nu} Z^{\mu \nu}
+\frac{1}{2} \frac{1}{1+\epsilon_2} (1+\frac{v^2}{2} C_{HD}) \frac{1}{4} (g^2+{g'}^2) v^2 Z_{\mu} Z^{\mu} .
\end{equation}
At this stage, we obtain a mass for the $Z_\mu$ field
\begin{equation}
M_Z=\sqrt{\frac{1+\frac{v^2}{2} C_{HD}}{1+\epsilon_2}} \sqrt{\frac{1}{4} (g^2+{g'}^2) v^2 }
\simeq \left(1+\frac{v^2}{4} C_{HD} -\frac{1}{2} \epsilon_2 \right) \sqrt{\frac{1}{4} (g^2+{g'}^2) v^2 } ,
\end{equation}
up to the linear order of the SMEFT operators.

Now we consider $W^{\pm}$ part.
We consider the following redefinition
\begin{equation}
W^{\pm}_{\mu} \to \frac{1}{\sqrt{1-2 v^2 C_{HW}}} W^{\pm}_{\mu} \simeq (1+v^2 C_{HW}) W^{\pm}_{\mu} .
\end{equation}
from which the W boson mass is obtained as 
\begin{equation}
M_W^2=\frac{1}{1- 2v^2 C_{HW}} \frac{g^2 v^2}{4} \simeq (1+2 v^2 C_{HW})\frac{g^2 v^2}{4} .
\end{equation}

\subsection{Interaction Lagrangian for fermions}

The interaction Lagrangian of mass dimension four is given by
\begin{align}
\mathcal{L}_{\text{dim-4}}^{\rm fermion}
&=\bar{L} i \gamma^{\mu} \left(\partial_{\mu}+i\frac{g}{2} \tau^I W_{\mu}^I-i\frac{{g'}}{2} B_{\mu} \right) L
+\bar{R} i \gamma^{\mu} (\partial_{\mu}-i {g'} B_{\mu}) R \nonumber \\ 
&=({\text{kin.\ term}}) \nonumber \\ 
&\quad{}+\frac{g {g'}}{\sqrt{g^2+{g'}^2}} \left(1-\frac{\epsilon_1}{2} \right) A_{\mu} \bar{\psi}_L \gamma^{\mu} \psi_L 
+\frac{g {g'}}{\sqrt{g^2+{g'}^2}} \left(1-\frac{\epsilon_1}{2} \right) A_{\mu} \bar{\psi}_R \gamma^{\mu} \psi_R \nonumber \\ 
&\quad{}+\left[\frac{1}{2}  \frac{g^2-{g'}^2}{\sqrt{g^2+{g'}^2}} \left(1-\frac{\epsilon_2}{2} \right)+2\epsilon_3 \frac{g {g'}}{\sqrt{g^2+{g'}^2}} \right]
Z_{\mu}  \bar{\psi}_L \gamma^{\mu} \psi_L \nonumber \\ 
&\quad{}+\left[ -\frac{{g'}^2}{\sqrt{g^2+{g'}^2}} \left(1-\frac{\epsilon_2}{2} \right)+2\epsilon_3 \frac{g {g'}}{\sqrt{g^2+{g'}^2}} \right]
Z_{\mu}  \bar{\psi}_R \gamma^{\mu} \psi_R \nonumber \\
&\quad{}-\frac{g}{\sqrt{2}} (1+v^2 C_{HW}) (W_{\mu}^+ J^{\mu}+W_{\mu}^- {J^{\mu}}^{\dagger})+\cdots
\end{align}
Here $\psi$ denotes a charged lepton and $J_{\mu}=\bar{\nu}_L \gamma^{\mu} e_L=\bar{\nu}_e \gamma^{\mu} P_L e+\bar{\nu}_{\mu} \gamma^{\mu} P_L \mu$.

The following dimension-six operators 
\begin{align}
&Q_{H \ell}^{(1)}
=(H^{\dagger} i \overleftrightarrow{D}_{\mu} H) (\bar{L} \gamma^{\mu} L) 
=v M_Z  Z_{\mu} (\bar{\psi} \gamma^{\mu} P_- \psi)+\cdots , \nonumber \\ 
&Q_{H \ell}^{(3)}
=(H^{\dagger} i \overleftrightarrow{D}_{\mu}^I H) (\bar{L} \tau^I \gamma^{\mu} L)
=- v M_Z Z_{\mu} (\bar{\psi} \gamma^{\mu} P_- \psi)-\frac{g}{\sqrt{2}} v^2 (W_{\mu}^+ J^{\mu}+W_{\mu}^-J^{\mu \dagger})+\cdots , \nonumber \\ 
&Q_{He}=(H^{\dagger} i \overleftrightarrow{D}_{\mu} H) (\bar{R} \gamma^{\mu} R)
=v M_Z  Z_{\mu}(\bar{\psi} \gamma^{\mu} P_+\psi ) +\cdots ,
\end{align}
further modify the interaction terms. 
We assume flavor conservation and flavor universality for these operators.
We obtain 
\begin{align}
\mathcal{L}_{\rm SMEFT}^{\rm fermion}
&=({\text{kin.\ term}}) \nonumber \\ 
&\quad{}+\frac{g {g'}}{\sqrt{g^2+{g'}^2}} \left(1-\frac{\epsilon_1}{2} \right) A_{\mu} \bar{\psi}_L \gamma^{\mu} \psi_L 
+\frac{g {g'}}{\sqrt{g^2+{g'}^2}} \left(1-\frac{\epsilon_1}{2} \right) A_{\mu} \bar{\psi}_R \gamma^{\mu} \psi_R \nonumber \\ 
&\quad{}+
\left[\frac{1}{2}  \frac{g^2-{g'}^2}{\sqrt{g^2+{g'}^2}} \left(1-\frac{\epsilon_2}{2} \right)+2\epsilon_3 \frac{g {g'}}{\sqrt{g^2+{g'}^2}}
+\frac{1}{2} v M_Z (C_{H \ell}^{(1)}-C_{H \ell}^{(3)}) \right]
Z_{\mu}  \bar{\psi}_L \gamma^{\mu} \psi_L \nonumber \\ 
&\quad{}+
\left[ -\frac{{g'}^2}{\sqrt{g^2+{g'}^2}} \left(1-\frac{\epsilon_2}{2} \right)+2\epsilon_3 \frac{g {g'}}{\sqrt{g^2+{g'}^2}} 
+\frac{1}{2} v M_Z C_{H \mu} \right]
Z_{\mu}  \bar{\psi}_R \gamma^{\mu} \psi_R \nonumber \\ 
&\quad{}-\frac{g}{\sqrt{2}} (1+v^2 C_{H \ell}^{(3)}+v^2 C_{HW}) (W_{\mu}^+ J^{\mu}+W_{\mu}^- {J^{\mu}}^{\dagger})+\cdots. \label{fermionint}
\end{align}
From this, the electric coupling constant is obtained as
\begin{equation}
e=\frac{g {g'}}{\sqrt{g^2+{g'}^2}} \left(1-\frac{\epsilon_1}{2} \right) .
\end{equation}

In SMEFT, fermions interact also through four-fermion interactions.
The relevant ones to our calculations are given by
\begin{align}
&Q_{\substack{\ell \ell \\ prst}}=(\bar{\ell}_p \gamma_{\mu} \ell_r)
(\bar{\ell}_s \gamma^{\mu} \ell_t)  , \\
&Q_{\substack{\ell e \\ prst}}=(\bar{\ell}_p \gamma_{\mu} \ell_r) 
(\bar{e}_s \gamma^{\mu} e_t ) , \\
&Q_{\substack{e e \\ prst}}=(\bar{e}_p \gamma_{\mu} e_r)
(\bar{e}_s \gamma^{\mu} e_t)  .
\end{align}
Here $p, r, s, t$ are flavor indices.
The Lagrangian is given by 
\begin{equation}
\mathcal{L}_{\text{four-fermi}}=
\sum_{p,r,s,t} ( C_{\substack{\ell \ell \\ prst}} Q_{\substack{\ell \ell \\ prst}}
+C_{\substack{\ell e \\ prst}} Q_{\substack{\ell e \\ prst}}
+C_{\substack{ee \\ prst}} Q_{\substack{ee \\ prst}}) . \label{fourfermi}
\end{equation}
For this Lagrangian, we impose
\begin{equation}
C_{\substack{\ell \ell \\ e \mu \mu e}}=C_{\substack{\ell \ell \\ \mu  e e \mu}}  \label{Clldef}
\equiv C_{\ell \ell} .
\end{equation}
We list the quantities on which
we can give constraints via measurements of the scatterings. 
\begin{align}
& C_{\substack{\ell \ell \\ \mu \mu \mu \mu}}   ,\quad 
 C_{\ell \ell}''\equiv \frac{1}{2} (C_{\substack{\ell \ell \\ e e \mu \mu}}+C_{\substack{\ell \ell \\ \mu \mu e e}}), \quad
 C_{\substack{\ell e \\ \mu \mu \mu \mu}} , \quad
 C_{\substack{\ell e \\ e e \mu \mu}} , \quad
 C_{\substack{\ell e \\ \mu \mu e e}} ,\quad
 C_{\substack{e e \\ \mu \mu \mu \mu}} ,\notag\\
 &C_{e \mu} \equiv \frac{1}{4}
(C_{\substack{e e \\ \mu \mu e e}}+C_{\substack{e e \\  e e \mu \mu}}
+C_{\substack{e e \\ \mu  e e \mu}}+C_{\substack{e e \\  e \mu \mu e}}) .
\end{align}
Note a Fierz identity $(\gamma_{\mu} P_{s})_{\alpha \beta}
(\gamma^{\mu} P_{s})_{\gamma \delta}=-(\gamma_{\mu} P_{s})_{\alpha \delta}
(\gamma^{\mu} P_{s})_{\gamma \beta}$.
We assume flavor conservation for these operators. Therefore, $(p,r,s,t)=(\mu, e, \mu, e)$ or $(e, \mu, e, \mu)$ type operator does not exist.

Integrating out the $W$ field in eq.~\eqref{fermionint} yields the following four-fermion Lagrangian:
\begin{align}
\mathcal{L}
&=- \frac{g^2 (1+v^2 C_{H \ell}^{(3)}+v^2 C_{HW})^2}{2 M_W^2} J_{\mu}^{\dagger} J^{\mu} +\cdots \nonumber \\
&=- \frac{g^2 (1+v^2 C_{H \ell}^{(3)}+v^2 C_{HW})^2}{2 M_W^2}  
\left[ (\bar{\nu}_e \gamma_{\mu} P_L e) ( \bar{\mu} \gamma^{\mu}P_L \nu_{\mu}) +(\bar{e} \gamma_{\mu} P_L \nu_e) (\bar{\nu}_{\mu} \gamma^{\mu} P_L \mu)+\cdots \right]. \label{JmuJmu}
\end{align}
Together with the four-fermion Lagrangian and eq.~\eqref{Clldef},
we obtain
\begin{equation}
\frac{G_F}{\sqrt{2}}=\frac{1}{2 v^2} \left(1+2 v^2 C_{H \ell}^{(3)}-v^2 C_{\ell \ell} \right)    ,
\end{equation}
where $C_{HW}$ in eq.~\eqref{JmuJmu} has been cancelled by that in $M_W$.

We summarize the important relations.
\begin{align}
&4 \pi \alpha=\frac{g^2 {g'}^2}{g^2+{g'}^2} (1- \epsilon_1)  , \label{match1} \\ 
&M_Z^2=\frac{g^2+{g'}^2}{4} v^2 \left(1+\frac{v^2}{2} C_{HD} -\epsilon_2 \right), \label{match2} \\ 
&\frac{G_F}{\sqrt{2}}=\frac{1}{2 v^2} \left(1+2 v^2 C_{H \ell}^{(3)}-v^2 C_{\ell \ell} \right) . \label{match3}
\end{align}
The quantities of left-hand side are accurately measured in experiments.
We use the following values:
\begin{equation}
\alpha_{EW}=127.95^{-1}, \quad{} M_Z=91.1876~{\rm GeV}, \quad{} G_F=1.16638 \times 10^{-5}~{\rm GeV}^{-2} .
\end{equation}
The SM gauge couplings and SM Higgs VEV are determined by setting the dimension-six
operator contributions to zero, i.e.,
\begin{align}
&4 \pi \alpha=\frac{g_{\rm SM}^2 {g'}_{\rm SM}^2}{g_{\rm SM}^2+{g'}_{\rm SM}^2}   , \label{match1SM}\\
&M_Z^2=\frac{g_{\rm SM}^2+g_{\rm SM}'^2}{4} v_{\rm SM}^2  , \label{match2SM}\\
&\frac{G_F}{\sqrt{2}}=\frac{1}{2 v_{\rm SM}^2} . \label{match3SM}
\end{align}
Here is the (tree-level) {\it{definition}} of $\{{g'}_{\rm SM}, g_{\rm SM}, v_{\rm SM}\}$.
The correct relations when assuming SMEFT are given by eqs.~\eqref{match1} --\eqref{match3}, and hence we expand $g$, ${g'}$, and $v$
as $g=g_{\rm SM}+\delta g$, ${g'}={g'}_{\rm SM}+\delta {g'}$, 
and $v=v_{\rm SM}+\delta v$, where perturbative corrections $\delta ...$
are given by linear combinations of the coefficients of 
the dimension-six operators.
Once we obtain the gauge couplings and VEV in this manner,
SMEFT gives non-trivial predictions for physical observables except for $\{\alpha, M_Z^2, G_F \}$.
For instance, SMEFT predicts the $W$ boson mass as
\begin{align}
M_W^2
&=(1+2 v^2 C_{HW})\frac{g^2 v^2}{4} \\ \nonumber
&=[1+2 (v_{\rm SM}^2+\delta v^2) C_{HW}]\frac{(g_{\rm SM}+\delta g)^2 (v_{\rm SM}^2+\delta v^2)}{4} \nonumber \\
&=M_{W, {\rm SM}}^2 \bigg[
1- \frac{1}{2} \frac{\cos^2{\theta_{\rm SM}}}{\cos^2{\theta_{\rm SM}}-\sin^2{\theta_{\rm SM}}} v^2_{\rm SM} C_{HD}  \nonumber \\
&\qquad{}\qquad{}\quad{} -\frac{\sin^2{\theta_{\rm SM}}}{\cos^2{\theta_{\rm SM}}-\sin^2{\theta_{\rm SM}}} v^2_{\rm SM} \left(2 C_{H \ell}^{(3)}-C_{\ell \ell} \right)  
-\frac{2 \cos{\theta_{\rm SM}} \sin{\theta_{\rm SM}}  }{\cos^2{\theta_{\rm SM}}-\sin^2{\theta_{\rm SM}}} v^2_{\rm SM} C_{HWB}
\bigg] , \label{Wbosonmass}
\end{align}
where $M_{W, {\rm SM}}^2=g_{\rm SM}^2 v_{\rm SM}^2/4$, 
$\cos{\theta_{\rm SM}}=g_{\rm SM}/\sqrt{g_{\rm SM}^2+({g'}_{\rm SM})^2}$.
In this paper, we give SMEFT predictions for the elastic scatterings.
The calculations are based on the results for
$g=g_{\rm SM}+\delta g$, ${g'}={g'}_{\rm SM}+\delta {g'}$, 
and $v=v_{\rm SM}+\delta v$ and the Lagrangian eq.~\eqref{fermionint}
and eq.~\eqref{fourfermi}.

It is often convenient to parameterize contributions from new physics in terms of the oblique $S$ and $T$ parameters~\cite{Peskin:1990zt,Altarelli:1990zd}.
They are related with the SMEFT operator within our basis as
\begin{equation}
    \frac{v^2}{\Lambda^2} C_{HWB} = \frac{g' g}{16 \pi}S ,\quad 
    \frac{v^2}{\Lambda^2} C_{HD} = - \frac{g'^2 g^2}{2 \pi(g^2 + g'^2)}T,
\end{equation}
which enable us to translate the constraints on the SMEFT operators into those on the electroweak precision observables.

\section{Precision measurements at a \texorpdfstring{$\mu^+ \mu^+$}{mu+ mu+} collider}
\label{sec:3}

In this section, we calculate the process 
$\mu^+ \mu^+ \to \mu^+ \mu^+$ using the SMEFT Lagrangian
discussed above.
First, we give the SM amplitude. 
Let $s_1$ and $s_2$ be the polarizations of initial muons.
Then the magnitude of the amplitude is given by
\begin{align}
&\sum_{s_3,s_4} |\mathcal{M}_{s_1 s_2}^{\rm SM}|^2 \nonumber \\
&=\sum_{i,j=\gamma, Z}  g^i_{-s_1} g^j_{-s_1} g^i_{-s_2} g^j_{-s_2} \nonumber \\
&\quad{} \times \bigg\{8 [(1+s_1 s_2)(p_1 \cdot p_2) (p_3 \cdot p_4)+(1-s_1 s_2)(p_1 \cdot p_4) (p_2 \cdot p_3) ] \frac{1}{D_i(p_1-p_3)}  \frac{1}{D_j(p_1-p_3)} \nonumber \\
&\qquad{}\quad{} + 8[(1+s_1 s_2)(p_1 \cdot p_2) (p_3 \cdot p_4)+(1-s_1 s_2)(p_1 \cdot p_3) (p_2 \cdot p_4) ] \frac{1}{D_i(p_1-p_4)}  \frac{1}{D_j(p_1-p_4)} \nonumber \\
&\qquad{}\quad{}+16  \delta_{s_1 s_2}  (p_1 \cdot p_2) (p_3 \cdot p_4)  \left[ \frac{1}{D_i(p_1-p_3)} \frac{1}{D_j(p_1-p_4)}+\frac{1}{D_i(p_1-p_4)} \frac{1}{D_j(p_1-p_3)} \right] \bigg\} ,
\end{align}
where $D_i(p) \equiv p^2-M_i^2$ with $M_{\gamma}=0$ and $M_Z \neq 0$.
We summed up the spins of the final muons $s_3$ and $s_4$. 
We denote by $p_1$ and $p_2$ the initial muon momenta and
by $p_3$ and $p_4$ those of the final muons.
We have
\begin{equation}
p_1 \cdot p_2=p_3 \cdot p_4=\frac{s}{2}, \quad{}
p_1 \cdot p_4=p_2 \cdot p_3=\frac{s}{2}(1-y), \quad{}
p_1 \cdot p_3=p_2 \cdot p_4=\frac{s}{2} y
\end{equation}
with $y=(1-\cos{\theta})/2$.
The coupling $g^Z_+$ is defined by
\begin{equation}
\mathcal{L}=g^Z_+ Z^{\mu} \bar{\psi}  \gamma_{\mu} P_+ \psi+\cdots
\end{equation}
with $P_{\pm}=(1 \pm \gamma_5)/2$. The other couplings are
understood in a similar manner.
The total cross section is given by
\begin{equation}
\sigma=\frac{1}{64 \pi s} 
\int_{\theta_{\rm min}}^{\pi-\theta_{\rm min}} d \theta
\sin{\theta} \sum_{s_3,s_4} |\mathcal{M}_{s_1 s_2}|^2 .
\end{equation}
Note that the final particles are identical.
We introduced an cutoff to the angular integration.
(Otherwise, the total cross section diverges.)
Furthermore, we divide the angle range $[\theta_{\rm min} , \pi -\theta_{\rm min}]$ into some bins.

The calculation using the SMEFT Lagrangian
\eqref{fermionint} requires a slight modification to the SM calculation;
it is sufficient to shift the SM coupling constants appropriately.
In addition, we have to calculate the four-fermion 
interaction contribution.
We consider the interference between the SM and the four-fermion
contribution.

In the following analysis, we turn on one of 
the dimension-six operator coefficients
and study how well we can constrain it
through the scattering experiments. 
We repeat this kind of analysis
for all the coefficients.
Schematically we can give the cross section integrated over one bin (at $\theta=\theta_i$) as
\begin{equation}
\sigma(\theta_i)=\sigma_{\rm SM}(\theta_i)+\hat{C} \sigma_{\rm NP}(\theta_i) .
\end{equation}
Here $\sigma_{\rm SM}(\theta_i)$ represents the SM cross section,
while $\hat{C} \sigma_{\rm NP}(\theta_i)$ represents the contribution from
a focused dimension-six operator.
$\hat{C}$ is a dimensionless coefficient, $\hat{C}=C \times {\rm TeV}^2$,
originated from the dimension-six operator coefficient $C$.
The coefficient $\hat{C}$ is determined by a fit in actual experiments using the $\chi^2$-test.
Here, we give a constraint on it assuming that 
no deviation from SM predictions is observed.
Then $\chi^2$ is given by
\begin{equation}
\chi^2=\sum_{i: \text{bin}} 
\left[\frac{\hat{C} \sigma_{\rm NP}(\theta_i)}{\Delta \sigma(\theta_i)} \right]^2 ,    
\end{equation}
where the statistical error on the cross section is assumed to be
\begin{equation}
\Delta \sigma(\theta_i)=\frac{\sigma_{\rm SM}(\theta_i)}{\sqrt{\int \mathcal{L} dt  \cdot \sigma_{\rm SM}(\theta_i)}}    .
\end{equation}
A two-sigma constraint on $\hat{C}$ can be then obtained as
\begin{equation}
|\hat{C}| < \sqrt{\frac{4}{\sum_{i: \text{bin}} 
\left[\frac{\sigma_{\rm NP}(\theta_i)}{\Delta \sigma(\theta_i)} \right]^2}} .
\end{equation}
We give our constraints in terms of new physics (or cutoff) scales.
Namely, we define a new physics scale as
\begin{equation}
C := \frac{1}{\Lambda^2}     .
\end{equation}
Then we obtain the two-sigma constraint for the new physics scale as
\begin{equation}
\Lambda >    \left( \frac{4}{\sum_{i: \text{bin}} 
\left[\frac{\sigma_{\rm NP}(\theta_i)}{\Delta \sigma(\theta_i)} \right]^2} \right)^{-1/4}~{\rm TeV} .
\end{equation}

\begin{table}[t]
    \begin{center}
    \begin{tabular}{c|ccc}
                                & RR           & LL               & RL   \\ \hline
    $C_{HWB}$           & 10~TeV    &  9.4~TeV       &  2.3~TeV \\ 
    $C_{HD}$              & 5.5~TeV   &  3.5~TeV       &  2.3~TeV \\
    $C_{H\ell}^{(1)}$    & 8.0~TeV   &  0                &   4.9~TeV\\
    $C_{H\ell}^{(3)}$    & 14~TeV    &  7.0~TeV      &  6.7~TeV \\
    $C_{H e}$         & 0             &   7.5~TeV     &  5.3~TeV  \\
    $C_{\ell \ell}$         & 7.7~TeV &  5.0~TeV      &  3.3~TeV\\
    $C_{\substack{\ell \ell \\ \mu \mu \mu \mu}}$         & 100~TeV &  0      &  0\\
    $C_{\substack{ee \\ \mu \mu \mu \mu}}$     & 0             &   100~TeV    &   0 \\
    $C_{\substack{\ell e \\ \mu \mu \mu \mu}}$       & 0            &    0               &   46~TeV \\
     \hline
    \end{tabular}
    \caption{Constraints on SMEFT operators at 2-sigma level. $\sqrt{s}=2$~TeV.
    The bin size for $\theta$ is taken as $1^\circ$ and each bin covers the range $ \theta_i-0.5^{\circ} <\theta < \theta_i +0.5^{\circ}$. 
    The considered range of $\theta_i$ is $16^\circ \leq \theta_i \leq 164 ^\circ$.}
    \label{tab:mumures}
    \end{center}
    \end{table}

\begin{figure}[t]
\begin{minipage}[b]{0.45\hsize}
\centering
\includegraphics[width=0.9\hsize]{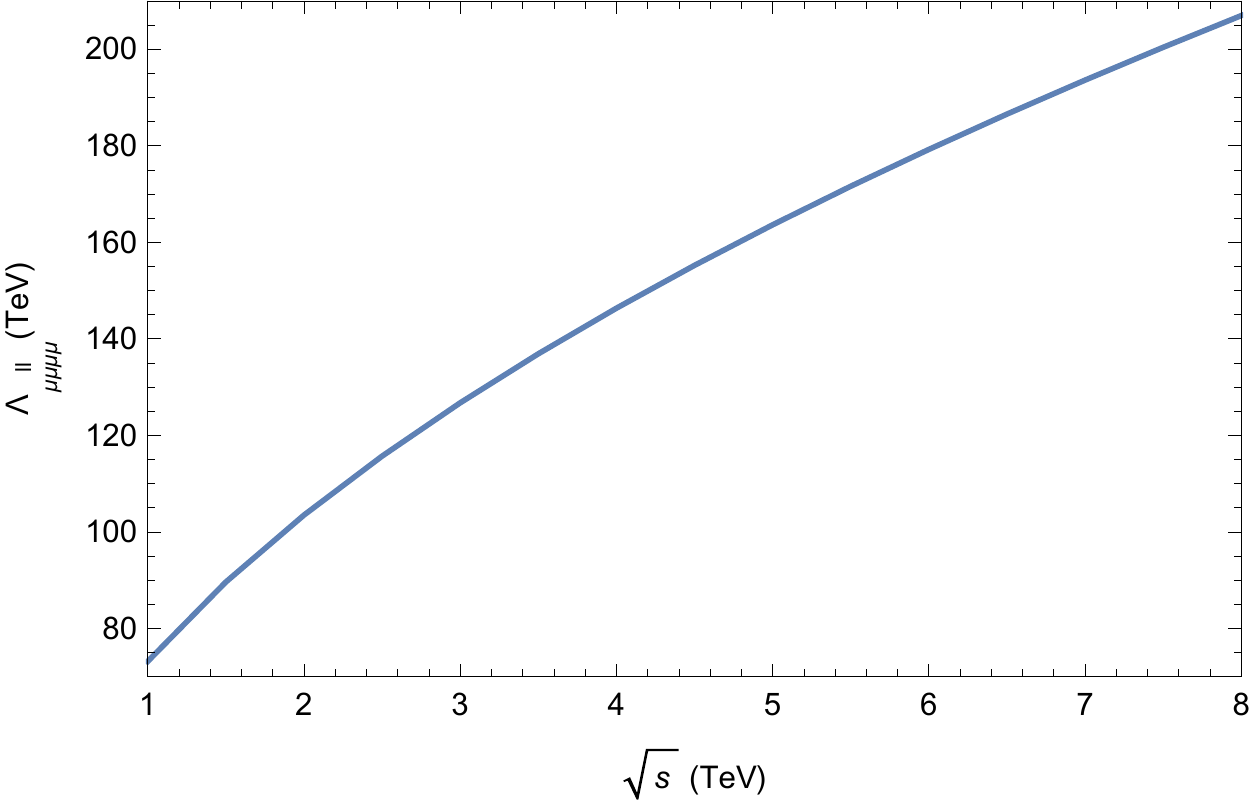}
\end{minipage}
\begin{minipage}[b]{0.45\hsize}
\centering
\includegraphics[width=0.9\hsize]{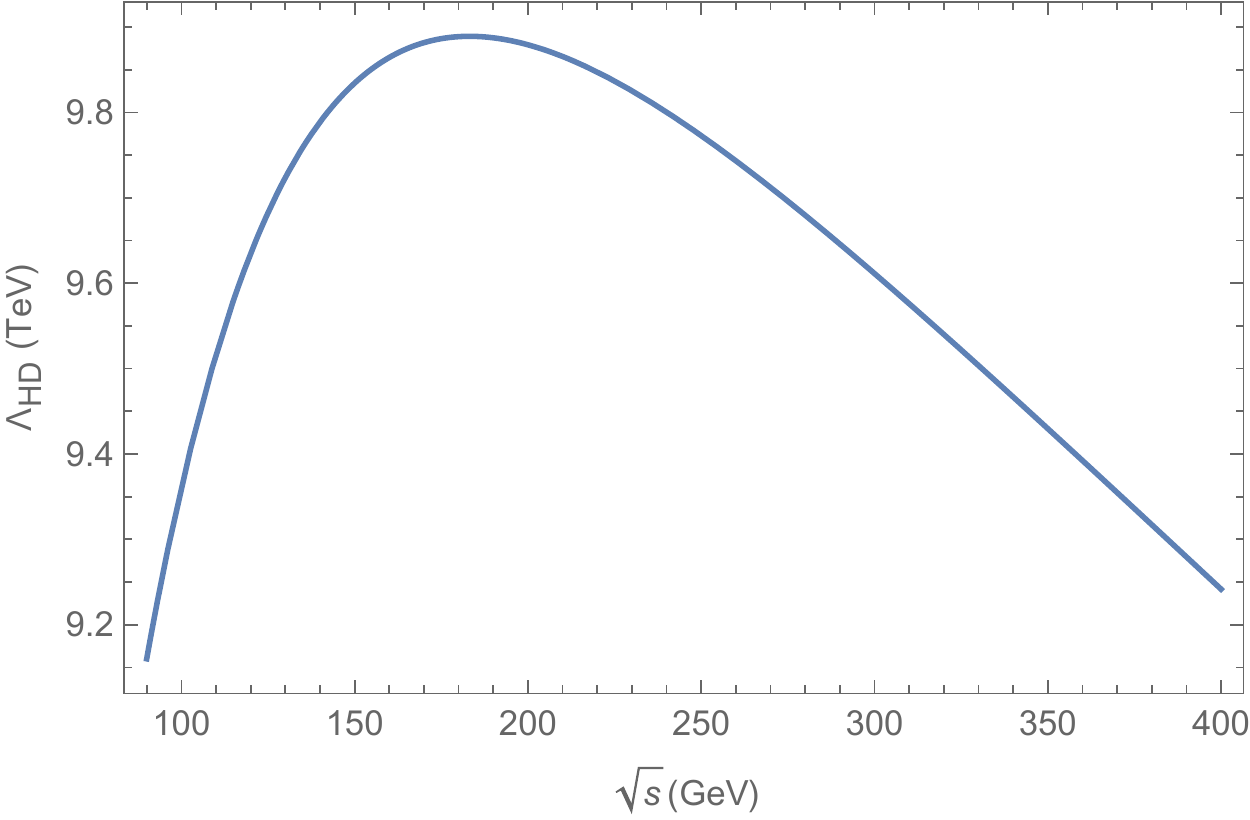}
\end{minipage}
\caption{The $\sqrt{s}$ dependence of the bounds on $\Lambda_{\ell \ell, \mu \mu \mu \mu}$ (left) and $\Lambda_{HD}$ (right) at the $\mu^+\mu^+$ collider. The initial helicity is set to be $s_1=s_2=1$.}
\label{fig:mumu_sdep}
\end{figure}

We consider the range $16^{\circ}<\theta < 164^{\circ}$ and take the
bin size as $1^{\circ}$. For instance, $\sigma(\theta_i=16^{\circ})$ means the
cross section integrated over $15.5^{\circ} < \theta < 16.5^{\circ}$.
The number of bins is 148. The constraint we expect to obtain is given
in tab.~\ref{tab:mumures} for $\sqrt{s}=2\ \text{TeV}$. We assume the
integrated luminosity to be $120~{\rm fb}^{-1}$. The initial helicity
corresponds to R: $s=+1$ and L: $s=-1$. We can obtain a powerful
constraint on four-fermion operators of $\mathcal{O}(100)$~TeV. The
$\sqrt{s}$ dependence of the bounds of $\Lambda_{\substack{\ell \ell \\ \mu \mu \mu \mu}}$ and
$\Lambda_{HD}$ is given in fig.~\ref{fig:mumu_sdep}.

\begin{figure}[t]
\begin{minipage}[b]{0.45\hsize}
\centering
\includegraphics[width=0.9\hsize]{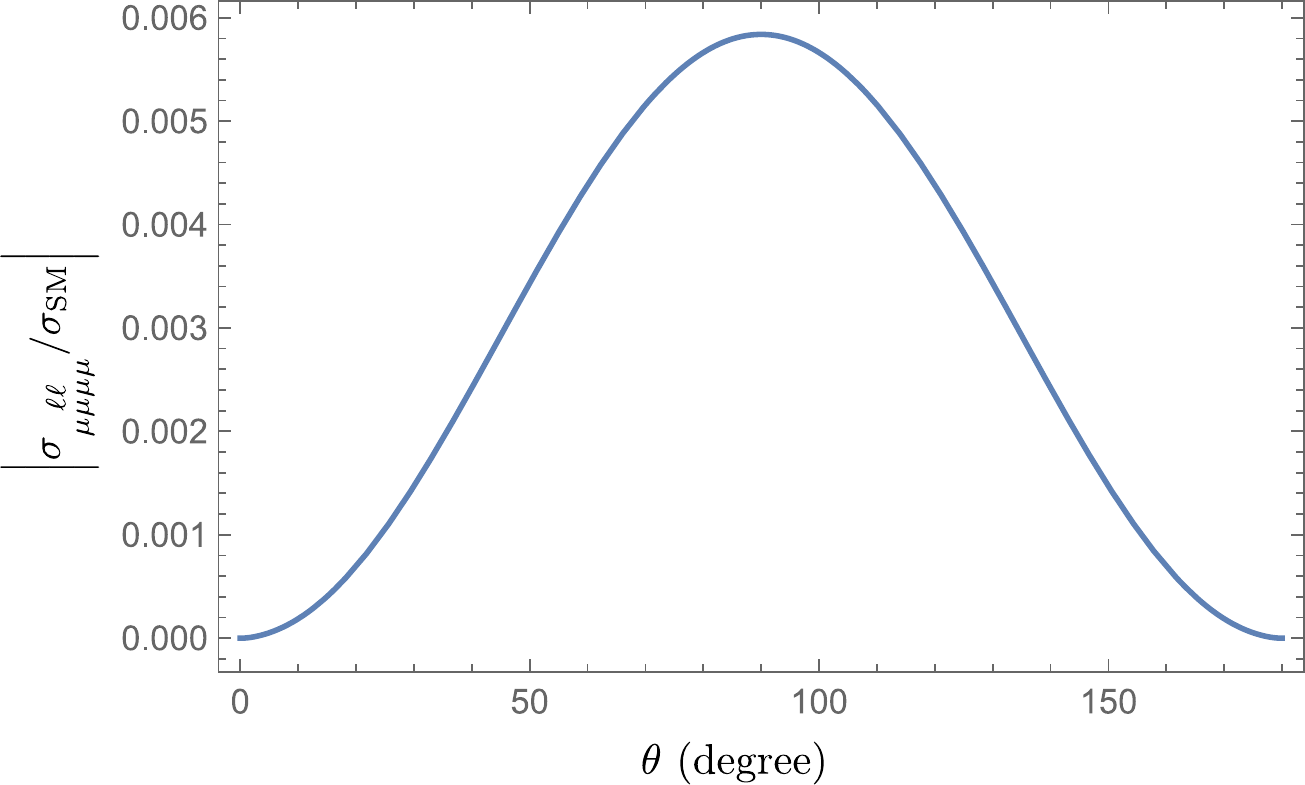}
\end{minipage}
\begin{minipage}[b]{0.45\hsize}
\centering
\includegraphics[width=0.9\hsize]{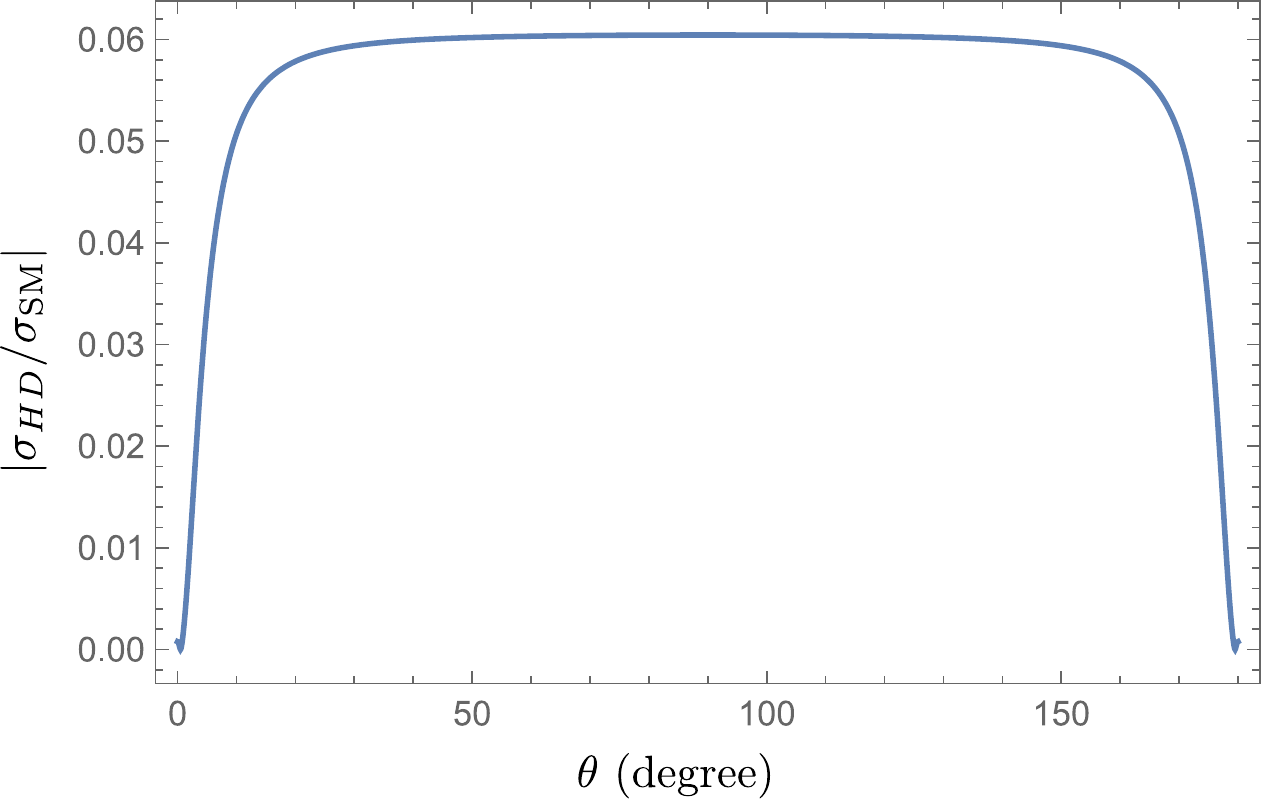}
\end{minipage}
\caption{The $\theta$ dependence of the ratio of the new physics contribution to the SM one for $C_{\ell\ell,\mu\mu\mu\mu}=1/(100\,\text{TeV})^2$ (left) and $C_{HD}=1/(\text{TeV}^2)$ (right) at the $\mu^+\mu^+$ collider of $\sqrt{s}=2\,\text{TeV}$. 
The initial helicity is set to be $s_1=s_2=1$.}
\label{fig:mumu_thetadep}
\end{figure}

We need to check whether the $\theta$ dependence of the new physics
contribution is different from the SM contribution, because the
luminosity would be measured by using 
the same scattering process. We
checked that the ratio of the new physics contribution to the SM one
has nontrivial dependence on $\theta$. See
fig.~\ref{fig:mumu_thetadep}.

\section{Precision measurements at an \texorpdfstring{$e^-\mu^+ $}{e- mu+} collider}
\label{sec:4}

Now we consider the scattering $e^- \mu^+ \to e^- \mu^+$.
The SM cross section is given through
\begin{align}
&\sum_{s_3, s_4} |\mathcal{M}_{s_1 s_2}^{\rm SM}|^2 \nonumber \\
&=\sum_{i, j} g^{i}_{s_1} g^j_{s_1} g^i_{-s_2} g^j_{-s_2} 8 [(1+s_1 s_2) (p_1 \cdot p_2) (p_3 \cdot p_4)+(1-s_1 s_2) (p_1 \cdot p_4) (p_2 \cdot p_3)]\notag\\
&\quad\times\frac{1}{(p_1-p_3)^2-M_i^2} \frac{1}{(p_1-p_3)^2-M_j^2}.
\end{align}
We label the initial state $e^-$ by $(p_1, s_1)$,
the initial state $\mu^+$ by $(p_2, s_2)$,
the final state $e^-$ by $(p_3, s_3)$ and 
the final state $\mu^+$ by $(p_4, s_4)$.

\begin{figure}[t]
  \centering
  \vspace{5mm}
  \includegraphics[width=0.75\hsize]{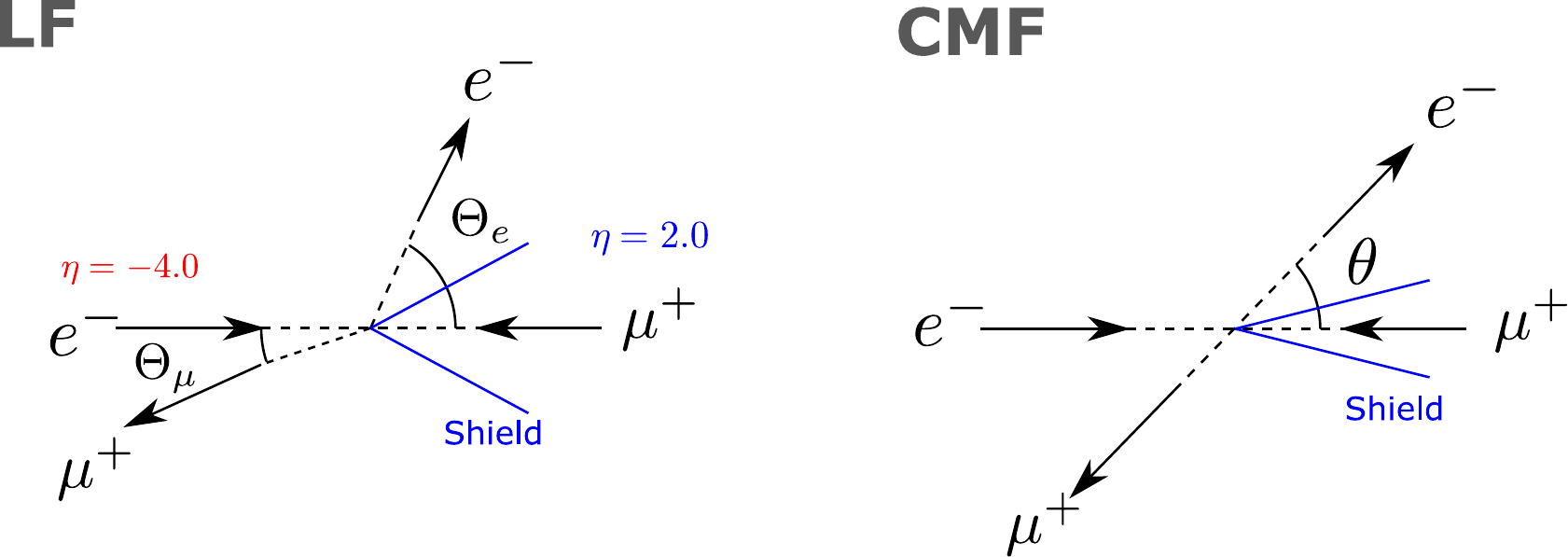}
  \caption{The angular range in which the electron and the muon go in the events we used. The left (right) figure represents the laboratory (center-of-mass) frame.}
  \label{fig:shield}
\end{figure}

\begin{figure}[t]
  \centering
  \includegraphics[width=0.5\hsize]{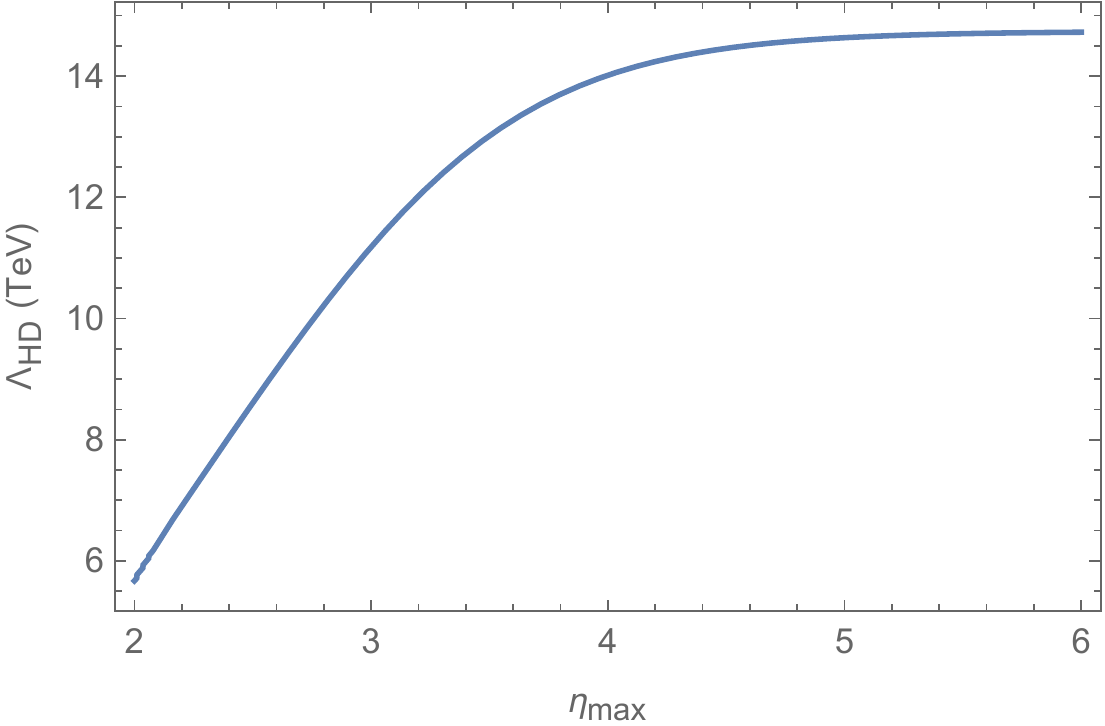}
  \caption{The $\eta$ cut (electron beam side) dependence of the bounds on $\Lambda_{HD}$ for the $e^-\mu^+$ collider of the initial energy $E_e=30\,\text{GeV}$ and $E_\mu=1\,\text{TeV}$.}
  \label{fig:emu_etamaxdep}
\end{figure}

In this analysis, we use the events where 
both an electron and a muon are observed in a certain range of angles.
We require both electron and muon go into the range of about
$15.4^{\circ}$ -- $178^{\circ}$ in the laboratory frame. This requirement
corresponds to setting the range of pseudo rapidity to be $-4 < \eta < 2$. (See fig.~\ref{fig:shield}.)
The asymmetric angular region is motivated by the fact that
we need to place a shield 
to protect the detector from decay products of the beam muons \cite{Bartosik:2020xwr}.
Although the detectability of the muons flying in the direction of the shield depends 
on the design of the detector, we take a conservative assumption that the particles flying into that angular region are not detected. 
However, since the produced particles tend to go in a direction 
of the electron beam side due to the beam energies,
this shield does not hinder us from catching events very much.
The practically important factor is how widely we can catch events on the electron beam side.
See fig.~\ref{fig:emu_etamaxdep} for the $\eta_{max}$ dependence of the bounds on $\Lambda_{HD}$.

In this analysis, we could obtain both the histograms concerning electrons and muons.
However, since we expect better angular resolution for electrons,
we only use the electron histogram.
As a result of the requirement mentioned above, 
the angular range is determined as $62.8^{\circ} \lesssim \Theta_e \lesssim 178^{\circ}$.
We summarize kinematic formulas 
in App.~\ref{app:A}.

\begin{figure}[t]
  \centering
  \includegraphics[width=0.5\hsize]{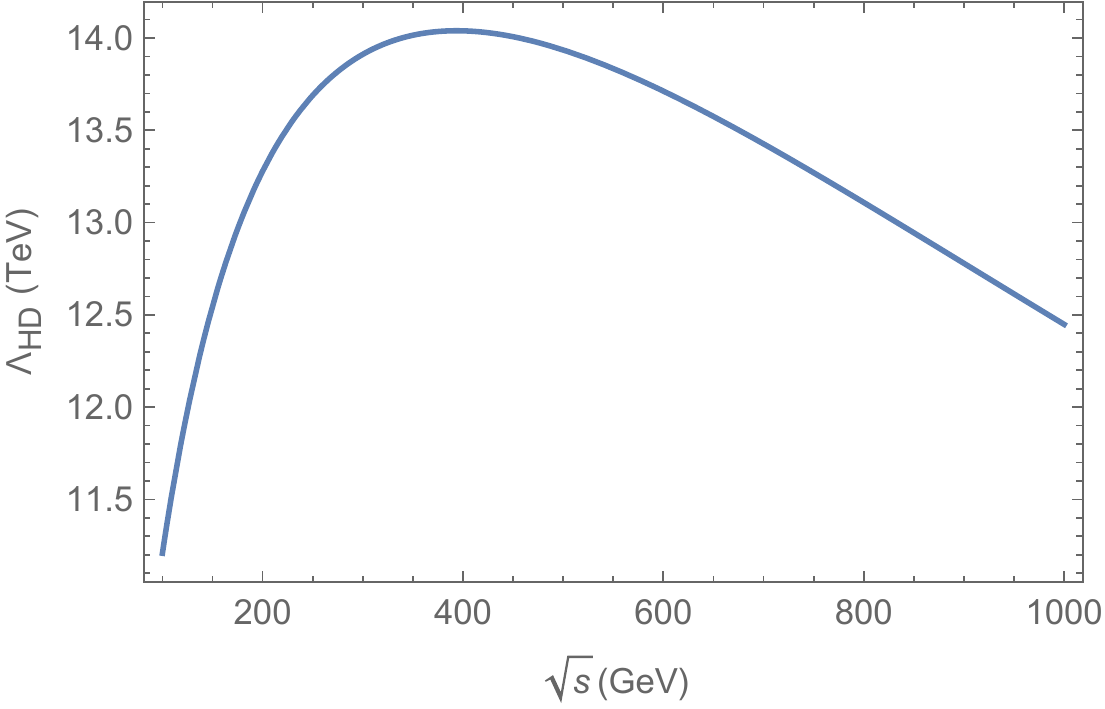}
\caption{The $\sqrt{s}$ dependence of the bounds on $\Lambda_{HD}$ for the $e^-\mu^+$ collider. The initial helicity is given as $s_1=-1,s_2=1$. We assume
the constant ratio 
$E_{\mu}/E_{e}=1000/30$
in drawing the figure.}
\label{fig:emu_rootsdep}
\end{figure}

\begin{figure}[t]
  \centering
  \includegraphics[width=0.5\hsize]{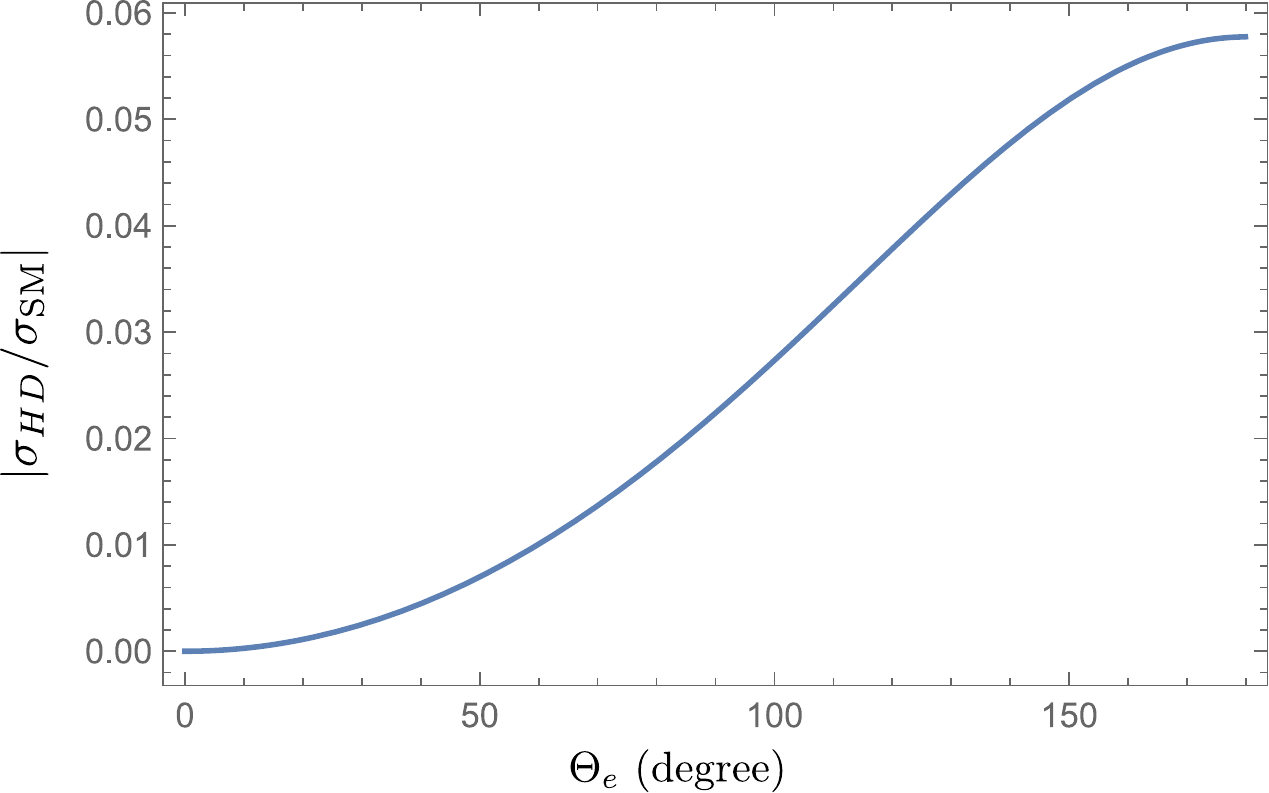}
  \caption{The $\Theta_e$ dependence of the ratio of the new physics contribution to the SM one for $C_{HD}=1/(\text{TeV}^{2})$ for the $e^-\mu^+$ collider. The initial helicity is given as $s_1=-1,s_2=1$.}
  \label{fig:emu_thetadep}
\end{figure}

The constraints we can obtain are summarized in tab.~\ref{tab:emures}.
We assume the integrated luminosity $\int \mathcal{L} dt=1~{\rm ab}^{-1}$.
\begin{table}[t]
\begin{center}
\begin{tabular}{c|cccc}
                            & RR           & RL               & LR                 & LL  \\ \hline
$C_{HWB}$           & 6.9~TeV    &  24~TeV       & 26~TeV      & 6.9~TeV \\ 
$C_{HD}$              & 6.8~TeV   &  9.0~TeV       &  14~TeV      & 6.8~TeV\\
$C_{H\ell}^{(1)}$    & 15~TeV   &  0                &   20~TeV       & 15~TeV \\
$C_{H\ell}^{(3)}$    & 20~TeV    &  18~TeV      &  35~TeV        & 20~TeV \\
$C_{H e}$         & 16~TeV    &  19~TeV     & 0                   & 16~TeV  \\
$C_{\ell \ell}$         & 9.6~TeV &  13~TeV      &  43~TeV           &  9.6~TeV  \\
$C''_{\ell \ell}$         & 0          &  0                 &  47~TeV           & 0 \\
$C_{e \mu}$     & 0             &   66~TeV    &       0                   & 0 \\
$C_{\substack{\ell e \\ e e \mu \mu }}$       & 0            &    0               &   0                  &44~TeV \\
$C_{\substack{\ell e \\ \mu \mu e e}}$       & 44~TeV            &    0               &   0            & 0  \\
 \hline
\end{tabular}
\caption{Constraints on SMEFT operators at two-sigma level. $E_e=30$~GeV
and $E_{\mu}=1$~TeV, which amounts to $\sqrt{s}=346$~GeV.
The bin size for $\Theta_e$ is taken as $1^\circ$. 
We require both muon and electron to go into the range of $15.4 ^\circ \lesssim \Theta \lesssim 178 ^\circ$,
corresponding to $\eta_{max}=2$ for the muon beam side and $\eta_{max}=4$ for the electron beam side.
As a result, the angle range of the electron is $62.8^{\circ} \lesssim \Theta_e \lesssim 178^{\circ}$.
}
\label{tab:emures}
\end{center}
\end{table}
It is worth noting that the constraints except for on four-fermion operators are 
stronger than at the $\mu^+ \mu^+$ collider.
The $\sqrt s$ dependence of the bounds and the $\theta$ dependence of the ratio of the new physics contribution to the SM, respectively, are given in figs.~\ref{fig:emu_rootsdep} and \ref{fig:emu_thetadep} for $C_{HD}$.

\section{Comparison to the current limits}
\label{sec:5}

\begin{figure}[ht]
  \centering
  \includegraphics[width=14cm]{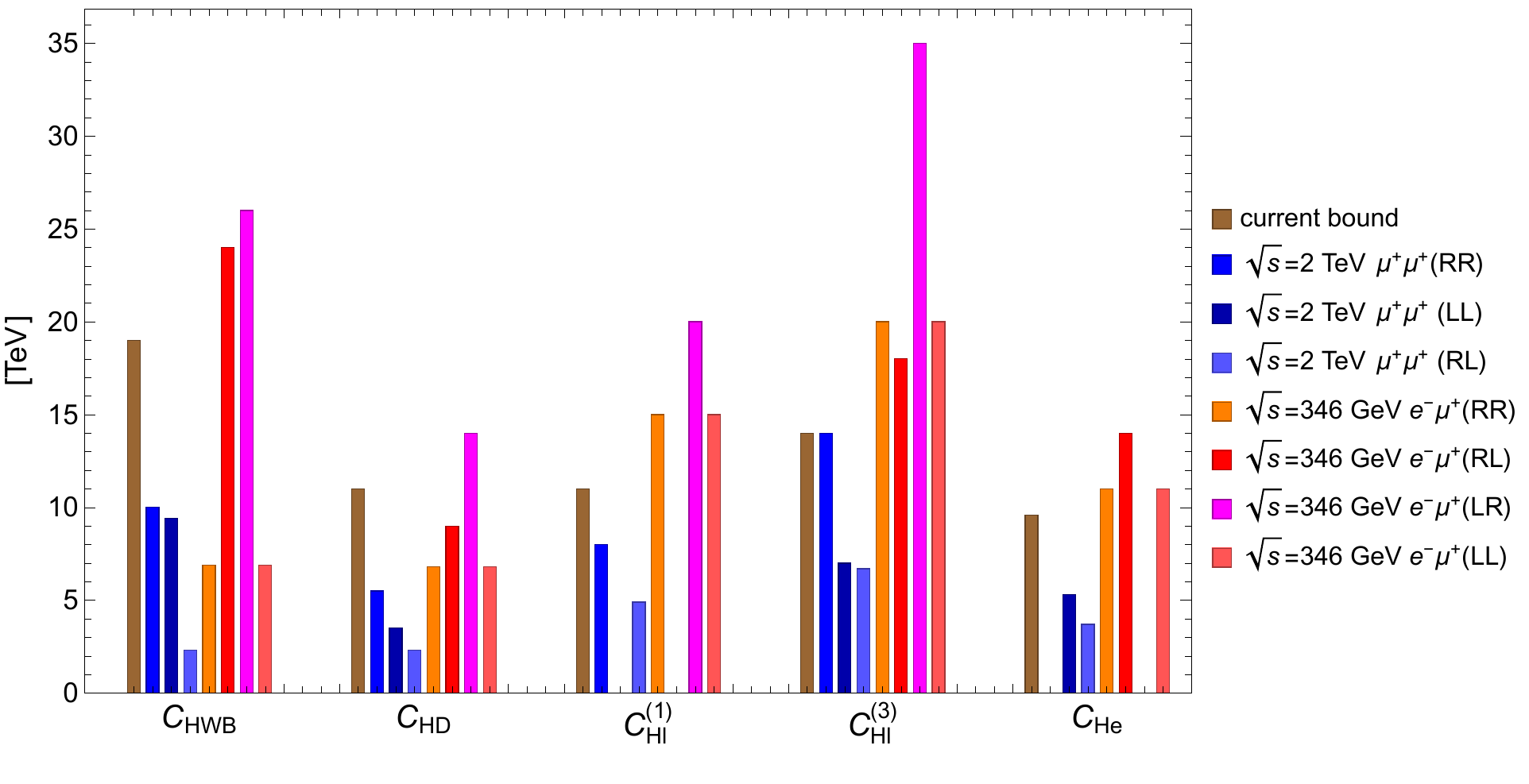}
  \caption{The current constraint and expected constraints from various scattering processes at $\mu$TRISTAN.}
  \label{fig:summary1}
\end{figure}

\begin{figure}[ht]
  \centering
  \includegraphics[width=14cm]{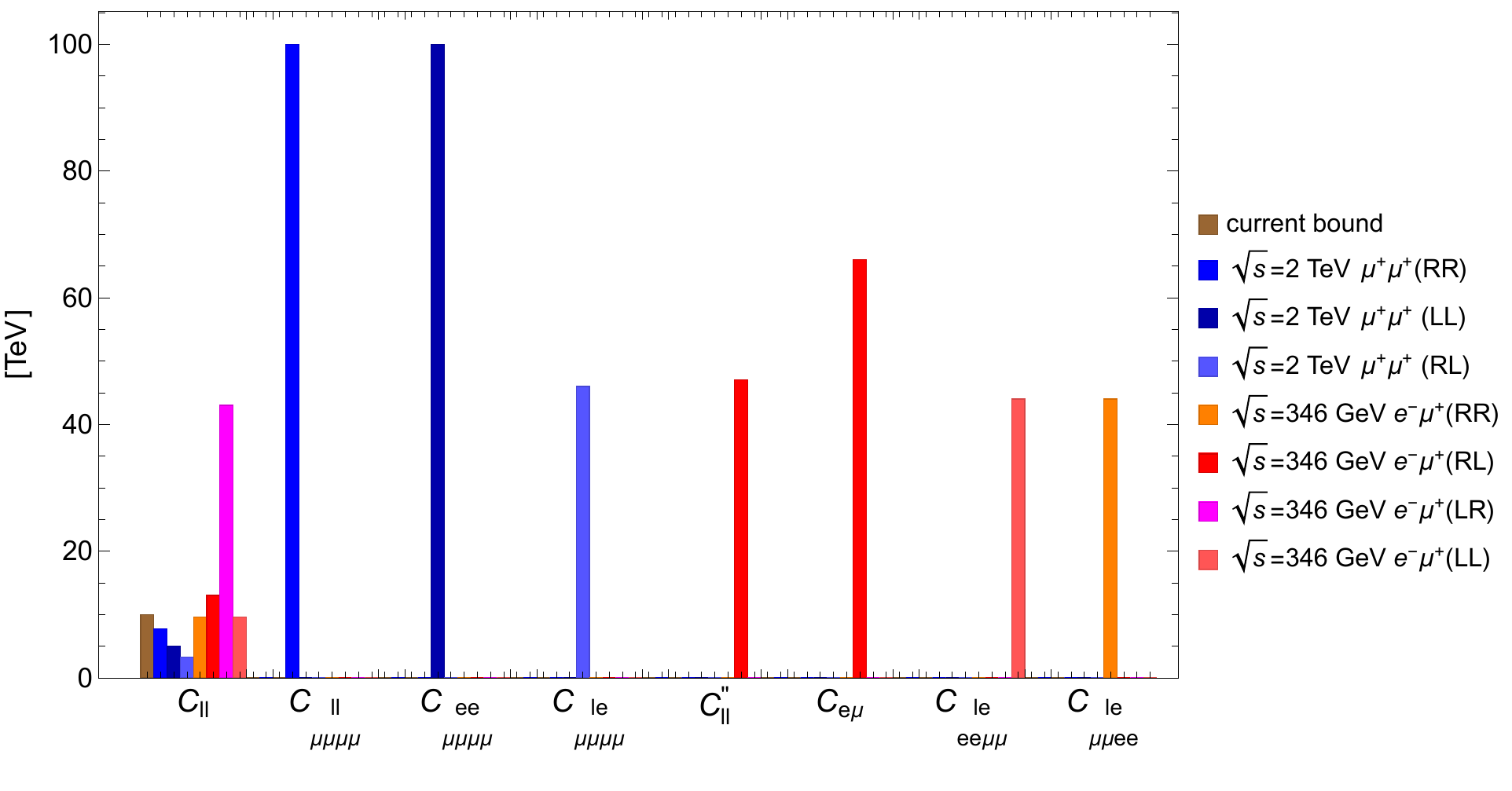}
  \caption{The current constraint and expected constraints from various scattering processes at $\mu$TRISTAN.}
  \label{fig:summary2}
\end{figure}

We compare the above expected constraints
with the current constraints. 
In Figs.~\ref{fig:summary1} and \ref{fig:summary2}, 
our results are compared with the current bounds given by Table 1 of 
ref.~\cite{Bagnaschi:2022whn}.\footnote{In Figs.~\ref{fig:summary1} and \ref{fig:summary2},
the current bound, for instance, of 19~{\rm TeV} for $C_{HWB}$ means that the current (two-sigma level) error size of $C_{HWB}$ is $\delta C_{HWB}=1/(19~{\rm TeV})^2$.
} Here their ``Individual''  result is referred.
Constraints on all the operators relevant to this study are expected to be improved. 
In particular, constraints on four-fermion 
interactions can be drastically improved.

Our study is relevant to studies which 
explain the recent $W$ boson mass anomaly
\cite{CDF:2022hxs} using SMEFT. 
Our study shows that the coefficients relevant to the $W$ boson mass formula eq.~\eqref{Wbosonmass} can be studied 
in the scattering processes.
In ref.~\cite{Bagnaschi:2022whn}, it is shown
that non-zero $C_{HD}$ (with the other coefficients set to zero) 
may explain the anomaly. 
In this scenario, $C_{HD}$ was determined as
$C_{HD}=-[0.035, 0.019]/({\rm TeV}^2)$
with the error $\delta C_{HD}=0.012/({\rm TeV}^2)$ at the two-sigma level. 
Our study shows that the error of $C_{HD}$
can be reduced to $\delta C_{HD} \simeq 0.005/({\rm TeV}^2)$. 
Therefore, this improvement can show more clearly
whether $C_{HD}$ deviates from zero or not.
(In addition, we may be able to measure $M_W$ directly at these new collider experiments.)


\section{Summary}
\label{sec:6}

Lepton colliders are known to be quite powerful for precision
measurements. There have been extensive discussions of $e^+ e^-$,
$\mu^+ \mu^-$, and $e^- e^-$ colliders as next generation colliders. In
this paper, we studied the $\mu^+$ based colliders as realistic muon
beams can possibly be achieved much earlier than $\mu^-$ by using the
technology of the ultra-cold muons.

We assumed that the energy of the $\mu^+$ beam to be 1~TeV which is
discussed as the design parameter of
$\mu$TRISTAN~\cite{Hamada:2022mua} with the main ring of circumference
of 3~km. In the $\mu^+ \mu^+$ collider option, we find that one can
probe the new physics interactions up to 100~TeV. This means that, for
example, effects of new gauge interactions of the muon can be seen up
to the symmetry breaking scale of $O(100)$~TeV.

We also studied the reach of the $\mu^+ e^-$ collider with the energy
of the electron to be 30~GeV. This option is motivated by the
measurements of the coupling of the Higgs boson, which is copiously
produced through the $W$ boson fusion process. We find that the
running with such an energy is also optimized for the electroweak
precision measurements, and one can improve the constraints for
$C_{HWB}$, $C_{HD}$, $C_{Hl}$ and $C_{He}$.

We worked at tree level for the
calculation of the SM processes.
This is sufficient to
approximately understand the
potential (best) reach of new 
physics scale.
However, it would be necessary to
sufficiently suppress systematic
uncertainties of SM predictions by
performing loop-level calculations
for actual analyses and for more
precise estimate of the reach.
See e.g. ref.~\cite{Bondarenko:2022kxq}
for recent development in the SM calculation.

The energy of the muon beam can be much higher for a larger ring. In
that case, the sensitivity to four-lepton operators gets much better.
For other operators involving the Higgs fields, the modification of
the Higgs coupling would be more important at high energy. In
addition, one can also hope to find new particles directly. Clearly,
the development of the muon acceleration technology will be a quite
important key for future particle physics.

\section*{Acknowledgements}
The work is supported by JSPS KAKENHI Grant Numbers JP19H00689~(RK, RM),
JP19K14711 (HT), JP21H01086~(RK), JP21J01117~(YH) and MEXT KAKENHI
Grant Number JP18H05542 (RK, HT).

\appendix
\numberwithin{equation}{section}
\setcounter{equation}{0}

\section{The laboratory frame coordinates for the \texorpdfstring{$\mu^+e^-$}{mu+ e-} collider}
\label{app:A}

In the $\mu^+ e^-$ collider, we have to note that
beam energies are asymmetric. We give some formulae 
concerning kinematics.
In the center-of-mass frame, the momenta are given by
\begin{equation}
p_1=\frac{\sqrt{s}}{2}
\begin{pmatrix}
1 \\ 0 \\ 0 \\ 1
\end{pmatrix},
\quad{}
p_2=\frac{\sqrt{s}}{2}
\begin{pmatrix}
1 \\ 0 \\ 0 \\ -1
\end{pmatrix} ,
\quad{}
p_3=\frac{\sqrt{s}}{2}
\begin{pmatrix}
1 \\ 0 \\ \sin{\theta} \\ \cos{\theta}
\end{pmatrix} ,
\quad{}
p_4=\frac{\sqrt{s}}{2}
\begin{pmatrix}
1 \\ 0 \\ -\sin{\theta} \\ -\cos{\theta}
\end{pmatrix} .
\end{equation}
The matrix for transforming to the laboratory frame is given by
\begin{equation}
M=
\begin{pmatrix}
\gamma & 0 & 0 & \beta \gamma \\
0 & 1 & 0 & 0 \\
0 & 0 & 1 & 0 \\
\beta \gamma & 0 & 0 & \gamma 
\end{pmatrix} .
\end{equation}
That is, the momenta in the laboratory frame is given by $p'_1=M p_1$ and so on.
The boost factor is obtained from
\begin{equation}
E_e=\frac{\sqrt{s}}{2} \gamma (1+\beta) , \quad{}
E_{\mu}=\frac{\sqrt{s}}{2} \gamma (1-\beta) ,
\end{equation}
where $\beta$ is explicitly given by
\begin{equation}
\beta=\frac{1-\frac{4 E_{\mu}^2}{s}}{1+\frac{4 E_{\mu}^2}{s}}
=\frac{E_e-E_{\mu}}{E_e+E_{\mu}} .
\end{equation}
The angle $\Theta_{\mu}$, in which the final muon goes in the laboratory frame, is given by
\begin{equation}
-\cos{\Theta}_{\mu}=\frac{p_{4,z}'}{|\vec{p}'_4|}
=\frac{\gamma(\beta-\cos{\theta})}{\sqrt{\sin^2{\theta}+\gamma^2 (\beta-\cos{\theta})^2}} ,
\end{equation}
which reads
\begin{equation}
\cos{\theta}=\frac{- \beta \gamma^2 \sin^2{\Theta_{\mu}} -\cos{\Theta_{\mu}}}{-\gamma^2+\cos{\Theta_{\mu}}^2(-1+\gamma^2)} .
\end{equation}
Then we have
\begin{equation}
d y=d (\frac{1-\cos{\theta}}{2})=-\frac{d (\cos{\theta})}{2}
=-\frac{1}{2}\frac{\gamma^2-2 \beta \gamma^2 \cos{\Theta_{\mu}}+(\gamma^2-1) \cos{\Theta_{\mu}}^2 }{(\gamma^2-(\gamma^2-1) \cos{\Theta_{\mu}}^2)^2} d (\cos{\Theta_{\mu}}) .
\end{equation}
The angle of the final electron in the laboratory frame, $\Theta_e$, is given similarly as
\begin{equation}
\cos{\Theta_e}=\frac{p_{3,z}'}{|\vec{p}'_3|}
=\frac{\gamma (\beta + \cos{\theta})}{\sqrt{\sin^2{\theta}+\gamma^2(\beta + \cos{\theta})^2 }} ,
\end{equation}
from which the following relations can be obtained:
\begin{equation}
\cos{\theta}=\frac{\beta \gamma^2 \sin^2{\Theta_e} - \cos{\Theta_e}}{-\gamma^2+\cos{\Theta_e}^2 (-1+\gamma^2)} ,
\end{equation}
\begin{equation}
dy =-\frac{1}{2}\frac{\gamma^2+2 \beta \gamma^2 \cos{\Theta_{e}}+(\gamma^2-1) \cos{\Theta_{e}}^2 }{(\gamma^2-(\gamma^2-1) \cos{\Theta_{e}}^2)^2} d (\cos{\Theta_{e}})  .
\end{equation}

\bibliography{bibcollection}

\begin{thebibliography}{10}

\bibitem{ALEPH:2005ab}
S.~Schael {\em et~al.}, {\it {Precision electroweak measurements on the $Z$
  resonance},} {\em Phys. Rept.}, vol.~427, pp.~257--454, 2006,
  \href{http://dx.doi.org/10.1016/j.physrep.2005.12.006}{{doi:10.1016/j.physrep.2005.12.006}},
   \href{http://arxiv.org/abs/hep-ex/0509008}{{\ttfamily
  arXiv:hep-ex/0509008}}.

\bibitem{ALEPH:2010aa}
{\it {Precision Electroweak Measurements and Constraints on the Standard
  Model},} 12 2010,  \href{http://arxiv.org/abs/1012.2367}{{\ttfamily
  arXiv:1012.2367\,[hep-ex]}}.

\bibitem{ALEPH:2013dgf}
S.~Schael {\em et~al.}, {\it {Electroweak Measurements in Electron-Positron
  Collisions at W-Boson-Pair Energies at LEP},} {\em Phys. Rept.}, vol.~532,
  pp.~119--244, 2013,
  \href{http://dx.doi.org/10.1016/j.physrep.2013.07.004}{{doi:10.1016/j.physrep.2013.07.004}},
   \href{http://arxiv.org/abs/1302.3415}{{\ttfamily
  arXiv:1302.3415\,[hep-ex]}}.

\bibitem{Derman:1979zc}
E.~Derman and W.~J. Marciano, {\it {Parity Violating Asymmetries in Polarized
  Electron Scattering},} {\em Annals Phys.}, vol.~121, p.~147, 1979,
  \href{http://dx.doi.org/10.1016/0003-4916(79)90095-2}{{doi:10.1016/0003-4916(79)90095-2}}.

\bibitem{Czarnecki:1995fw}
A.~Czarnecki and W.~J. Marciano, {\it {Electroweak radiative corrections to
  polarized Moller scattering asymmetries},} {\em Phys. Rev. D}, vol.~53,
  pp.~1066--1072, 1996,
  \href{http://dx.doi.org/10.1103/PhysRevD.53.1066}{{doi:10.1103/PhysRevD.53.1066}},
   \href{http://arxiv.org/abs/hep-ph/9507420}{{\ttfamily
  arXiv:hep-ph/9507420}}.

\bibitem{Ramsey-Musolf:1999qyv}
M.~J. Ramsey-Musolf, {\it {Low-energy parity violation and new physics},} {\em
  Phys. Rev. C}, vol.~60, p.~015501, 1999,
  \href{http://dx.doi.org/10.1103/PhysRevC.60.015501}{{doi:10.1103/PhysRevC.60.015501}},
   \href{http://arxiv.org/abs/hep-ph/9903264}{{\ttfamily
  arXiv:hep-ph/9903264}}.

\bibitem{Czarnecki:2000ic}
A.~Czarnecki and W.~J. Marciano, {\it {Polarized Moller scattering
  asymmetries},} {\em Int. J. Mod. Phys. A}, vol.~15, pp.~2365--2376, 2000,
  \href{http://dx.doi.org/10.1016/S0217-751X(00)00243-0}{{doi:10.1016/S0217-751X(00)00243-0}},
   \href{http://arxiv.org/abs/hep-ph/0003049}{{\ttfamily
  arXiv:hep-ph/0003049}}.

\bibitem{SLACE158:2005uay}
P.~L. Anthony {\em et~al.}, {\it {Precision measurement of the weak mixing
  angle in Moller scattering},} {\em Phys. Rev. Lett.}, vol.~95, p.~081601,
  2005,
  \href{http://dx.doi.org/10.1103/PhysRevLett.95.081601}{{doi:10.1103/PhysRevLett.95.081601}},
   \href{http://arxiv.org/abs/hep-ex/0504049}{{\ttfamily
  arXiv:hep-ex/0504049}}.

\bibitem{Kumar:2013yoa}
K.~S. Kumar, S.~Mantry, W.~J. Marciano, and P.~A. Souder, {\it {Low Energy
  Measurements of the Weak Mixing Angle},} {\em Ann. Rev. Nucl. Part. Sci.},
  vol.~63, pp.~237--267, 2013,
  \href{http://dx.doi.org/10.1146/annurev-nucl-102212-170556}{{doi:10.1146/annurev-nucl-102212-170556}},
   \href{http://arxiv.org/abs/1302.6263}{{\ttfamily
  arXiv:1302.6263\,[hep-ex]}}.

\bibitem{MOLLER:2014iki}
J.~Benesch {\em et~al.}, {\it {The MOLLER Experiment: An Ultra-Precise
  Measurement of the Weak Mixing Angle Using M{\textbackslash{}o}ller
  Scattering},} 11 2014,  \href{http://arxiv.org/abs/1411.4088}{{\ttfamily
  arXiv:1411.4088\,[nucl-ex]}}.

\bibitem{Buchmuller:1985jz}
W.~Buchmuller and D.~Wyler, {\it {Effective Lagrangian Analysis of New
  Interactions and Flavor Conservation},} {\em Nucl. Phys. B}, vol.~268,
  pp.~621--653, 1986,
  \href{http://dx.doi.org/10.1016/0550-3213(86)90262-2}{{doi:10.1016/0550-3213(86)90262-2}}.

\bibitem{Grzadkowski:2010es}
B.~Grzadkowski, M.~Iskrzynski, M.~Misiak, and J.~Rosiek, {\it {Dimension-Six
  Terms in the Standard Model Lagrangian},} {\em JHEP}, vol.~10, p.~085, 2010,
  \href{http://dx.doi.org/10.1007/JHEP10(2010)085}{{doi:10.1007/JHEP10(2010)085}},
   \href{http://arxiv.org/abs/1008.4884}{{\ttfamily
  arXiv:1008.4884\,[hep-ph]}}.

\bibitem{Brivio:2017vri}
I.~Brivio and M.~Trott, {\it {The Standard Model as an Effective Field
  Theory},} {\em Phys. Rept.}, vol.~793, pp.~1--98, 2019,
  \href{http://dx.doi.org/10.1016/j.physrep.2018.11.002}{{doi:10.1016/j.physrep.2018.11.002}},
   \href{http://arxiv.org/abs/1706.08945}{{\ttfamily
  arXiv:1706.08945\,[hep-ph]}}.

\bibitem{Ellis:2020unq}
J.~Ellis, M.~Madigan, K.~Mimasu, V.~Sanz, and T.~You, {\it {Top, Higgs, Diboson
  and Electroweak Fit to the Standard Model Effective Field Theory},} {\em
  JHEP}, vol.~04, p.~279, 2021,
  \href{http://dx.doi.org/10.1007/JHEP04(2021)279}{{doi:10.1007/JHEP04(2021)279}},
   \href{http://arxiv.org/abs/2012.02779}{{\ttfamily
  arXiv:2012.02779\,[hep-ph]}}.

\bibitem{Behnke:2013xla}
{\it {The International Linear Collider Technical Design Report - Volume 1:
  Executive Summary},} 6 2013,
  \href{http://arxiv.org/abs/1306.6327}{{\ttfamily
  arXiv:1306.6327\,[physics.acc-ph]}}.

\bibitem{Ellis:2015sca}
J.~Ellis and T.~You, {\it {Sensitivities of Prospective Future e+e- Colliders
  to Decoupled New Physics},} {\em JHEP}, vol.~03, p.~089, 2016,
  \href{http://dx.doi.org/10.1007/JHEP03(2016)089}{{doi:10.1007/JHEP03(2016)089}},
   \href{http://arxiv.org/abs/1510.04561}{{\ttfamily
  arXiv:1510.04561\,[hep-ph]}}.

\bibitem{Hamada:2022mua}
Y.~Hamada, R.~Kitano, R.~Matsudo, H.~Takaura, and M.~Yoshida, {\it
  {$\mu$TRISTAN},} {\em PTEP}, vol.~2022, no.~5, p.~053B02, 2022,
  \href{http://dx.doi.org/10.1093/ptep/ptac059}{{doi:10.1093/ptep/ptac059}},
  \href{http://arxiv.org/abs/2201.06664}{{\ttfamily
  arXiv:2201.06664\,[hep-ph]}}.

\bibitem{Kondo:2018rzx}
Y.~Kondo {\em et~al.}, {\it {Re-Acceleration of Ultra Cold Muon in J-PARC Muon
  Facility},} in {\em {9th International Particle Accelerator Conference}}, 6
  2018.

\bibitem{Abe:2019thb}
M.~Abe {\em et~al.}, {\it {A New Approach for Measuring the Muon Anomalous
  Magnetic Moment and Electric Dipole Moment},} {\em PTEP}, vol.~2019, no.~5,
  p.~053C02, 2019,
  \href{http://dx.doi.org/10.1093/ptep/ptz030}{{doi:10.1093/ptep/ptz030}},
  \href{http://arxiv.org/abs/1901.03047}{{\ttfamily
  arXiv:1901.03047\,[physics.ins-det]}}.

\bibitem{Peskin:1990zt}
M.~E. Peskin and T.~Takeuchi, {\it {A New constraint on a strongly interacting
  Higgs sector},} {\em Phys. Rev. Lett.}, vol.~65, pp.~964--967, 1990,
  \href{http://dx.doi.org/10.1103/PhysRevLett.65.964}{{doi:10.1103/PhysRevLett.65.964}}.

\bibitem{Altarelli:1990zd}
G.~Altarelli and R.~Barbieri, {\it {Vacuum polarization effects of new physics
  on electroweak processes},} {\em Phys. Lett. B}, vol.~253, pp.~161--167,
  1991,
  \href{http://dx.doi.org/10.1016/0370-2693(91)91378-9}{{doi:10.1016/0370-2693(91)91378-9}}.

\bibitem{Alonso:2013hga}
R.~Alonso, E.~E. Jenkins, A.~V. Manohar, and M.~Trott, {\it {Renormalization
  Group Evolution of the Standard Model Dimension Six Operators III: Gauge
  Coupling Dependence and Phenomenology},} {\em JHEP}, vol.~04, p.~159, 2014,
  \href{http://dx.doi.org/10.1007/JHEP04(2014)159}{{doi:10.1007/JHEP04(2014)159}},
   \href{http://arxiv.org/abs/1312.2014}{{\ttfamily
  arXiv:1312.2014\,[hep-ph]}}.

\bibitem{Bartosik:2020xwr}
N.~Bartosik {\em et~al.}, {\it {Detector and Physics Performance at a Muon
  Collider},} {\em JINST}, vol.~15, no.~05, p.~P05001, 2020,
  \href{http://dx.doi.org/10.1088/1748-0221/15/05/P05001}{{doi:10.1088/1748-0221/15/05/P05001}},
   \href{http://arxiv.org/abs/2001.04431}{{\ttfamily
  arXiv:2001.04431\,[hep-ex]}}.

\bibitem{Bagnaschi:2022whn}
E.~Bagnaschi, J.~Ellis, M.~Madigan, K.~Mimasu, V.~Sanz, and T.~You, {\it {SMEFT
  Analysis of $m_{W}$},} 4 2022,
  \href{http://arxiv.org/abs/2204.05260}{{\ttfamily
  arXiv:2204.05260\,[hep-ph]}}.

\bibitem{CDF:2022hxs}
T.~Aaltonen {\em et~al.}, {\it {High-precision measurement of the W boson mass
  with the CDF II detector},} {\em Science}, vol.~376, no.~6589, pp.~170--176,
  2022,
  \href{http://dx.doi.org/10.1126/science.abk1781}{{doi:10.1126/science.abk1781}}.

\bibitem{Bondarenko:2022kxq}
S.~G. Bondarenko, L.~V. Kalinovskaya, L.~A. Rumyantsev, and V.~L. Yermolchyk,
  {\it {One-loop electroweak radiative corrections to polarized M\o{}ller
  scattering},} 3 2022,  \href{http://arxiv.org/abs/2203.10538}{{\ttfamily
  arXiv:2203.10538\,[hep-ph]}}.

\end{thebibliography}
\bibliographystyle{arxiv-bibstyle/hyperieeetr2}

\end{document}